\documentclass{osa-article}

\journal{oe}

\usepackage{amsmath}

\DeclareMathOperator*{\argmin}{arg\,min}
\usepackage{mathtools}
\DeclarePairedDelimiter\abs{\lvert}{\rvert}
\DeclarePairedDelimiter\norm{\lVert}{\rVert}
\makeatletter
\let\oldabs\abs
\def\abs{\@ifstar{\oldabs}{\oldabs*}}
\let\oldnorm\norm
\def\norm{\@ifstar{\oldnorm}{\oldnorm*}}
\makeatother

\begin{document}

\title{Regularized Inverse Holographic Volume Reconstruction for 3D Particle Tracking}

\author{Kevin Mallery and Jiarong Hong\authormark{*}}

\address{Department of Mechanical Engineering, University of Minnesota, 111 Church St SE, Minneapolis, MN 55455, USA}

\email{\authormark{*}jhong@umn.edu} %

\begin{abstract}
The key limitations of digital inline holography (DIH) for particle tracking applications are poor longitudinal resolution, particle concentration limits, and case-specific processing. We utilize an inverse problem method with fused lasso regularization to perform full volumetric reconstructions of particle fields. By exploiting data sparsity in the solution and utilizing GPU processing, we dramatically reduce the computational cost usually associated with inverse reconstruction approaches. We demonstrate the accuracy of the proposed method using synthetic and experimental holograms. Finally, we present two practical applications (high concentration microorganism swimming and microfiber rotation) to extend the capabilities of DIH beyond what was possible using prior methods.
\end{abstract}

\section{Introduction}
Digital inline holography (DIH) is a diffractive imaging method in which volumetric information is encoded and subsequently extracted from a 2D image \cite{Poon2014}. 
The ability to resolve the position of objects in 3D naturally leads to temporal tracking \cite{Katz2010, Yu2014} with applications in particle dynamics \cite{Seifi2013}, microorganism swimming \cite{Molaei2014}, measurements of 3D flow fields \cite{Sheng2009}, and flow-structure interaction \cite{Zhang2015, Toloui2019} among many others. The primary advantage of DIH particle tracking velocimetry (DIH-PTV) is that it is able to capture time-resolved 3D velocities using only a single camera. 
DIH-PTV is substantially less expensive than methods such as tomographic PTV or traditional PIV due to the need for only one camera and compatibility with low cost laser sources. The low cost and hardware simplicity of DIH has enabled multiple in situ applications \cite{Beals2015a, Ozcan2016, Wu2017}.

Despite these advantages, there are several drawbacks of DIH-PTV that have limited broad application of the method. Since the inception of DIH-PTV, poor longitudinal (i.e. out of the plane of the image) resolution has consistently been the greatest challenge \cite{Hinsch2002, Katz2010, Toloui2017}. Known as the depth-of-focus (DOF) problem, the apparent elongation of reconstructed particles is caused by the finite sensor extent and the intensity integration effect of physical pixels \cite{Katz2010, Poon2014}. Additional imaging noise further reduces the ability to resolve the critical high frequency fringes. Because of this limitation, some applications of DIH-PTV (namely \cite{Ling2016, Toloui2017}) have primarily focused on the analysis of the much more accurate in-plane velocities. 
When high particle concentrations are used, cross-interference and other noise sources (twin image, out-of-focus particles) result in a low signal-to-noise ratio (SNR) \cite{Malek2004, Chen2015}. Malek et al. showed that the reconstruction quality depends on both the shadow density and the depth of the sample \cite{Malek2004}. The shadow density is
\begin{equation}
\label{eqn:shadow_density}
    s_d=n_sLd^2
\end{equation}
where \(n_s\) is particle number concentration, \(L\) is sample depth, and \(d\) is the particle diameter.
Methods for improving DIH-PTV (see next paragraph) often require increasing the optical complexity, extensive process tuning by an expert, and expensive computations.

Many approaches to overcoming these drawbacks focus on hardware design to improve the recorded image quality or encode more information in the recording. The most common is the use of multiple viewing angles (tomographic DIH), using the lateral accuracy and some depth information from each view to more accurately localize the particles \cite{Zhang1997, Kebbel1999, Buchmann2013, Gao2018}. This method requires only two cameras (or one by using mirrors \cite{Kebbel1999}) compared to four required for conventional tomography. Other approaches specific to DIH-PTV seek to reduce the effective shadow density by illuminating only a limited volume of interest \cite{Cao2008, Allano2013} or using localized particle seeding \cite{Talapatra2013}. Due to their mechanical and optical complexity, these methods are non-trivial to implement and care must be taken to avoid flow disturbance. Off-axis holography is also commonly used as it does not have the twin image problem and separates the reference beam from the object (reducing contamination) \cite{Lindensmith2016, Kuhn2014}. However, this requires precision alignment and higher laser coherence. Collectively, these methods requiring multiple optical paths, viewing angles, and calibration thereof negate the principle advantages of DIH-PTV, namely ease of use and hardware simplicity.

Many authors have focused on improving the numerical processing of DIH-PTV images. Much of this work has been focused on automatic detection of the object focal plane \cite{Burns2007, Wilson2012}. However, these do not address the problem of low SNR and usually assume that accurate 2D segmentation is trivial which is only true when the particle concentration is very low. Iterative phase retrieval methods have been shown to solve the twin image problem and improve the reconstructed SNR \cite{Denis2005, Latychevskaia2007}, but have not been applied for PTV. Holographic deconvolution \cite{Dixon2011, Latychevskaia2010, Latychevskaia2014, Toloui2015, Toloui2017} borrows a method from optical microscopy to treat the apparent blurring of point objects in the 3D reconstruction as convolution of the true object with a blurring point spread function (PSF).
However, the dependence of deconvolution on a 3D Fourier transform makes this method memory intensive and windowing may be needed to process large holograms (more than \(10^8\) voxels). The point object assumption also limits the range of applications suitable for deconvolution.

A more recent approach to holographic reconstruction is the inverse method \cite{Soulez2007a, Brady2009}. Inverse methods, rather than reconstructing the object from the image, instead find the optimal object that would produce the observed image while satisfying some physical constraints.
One of the first inverse problem formulations was proposed by Soulez et al. \cite{Soulez2007a} who performed a 4D parametric optimization to find the 3D location and radius of spherical particles.
Their stated linear time dependence on the number of particles makes this method unsuitable for fluid flow applications where thousands of particles must be tracked for hundreds of frames. Furthermore, the assumption of spherical particles restricts the scope of potential applications.
The use of the term "compressive holography" (CH) to refer to the inverse problem was introduced by Brady et al. in 2009 who borrowed concepts from the field of compressive sensing for holographic processing \cite{Brady2009}. They used a total variation regularized approach to produce in-focus images of two dandelion seed parachutes recorded concurrently at two different focal planes. Denis et al. \cite{Denis2009} used a similar approach with a simpler sparsity-based regularization. Both approaches show a significant reduction in the out-of-focus noise, twin images, and other noise. More recently, a set of physically meaningful constraints (including sparsity, smoothness, and non-amplification) have been used to achieve excellent reconstructions of absorption and phase of individual evaporating particles and their evaporation tails \cite{Jolivet2018}.

Inverse methods have only been used to reconstruct objects for which the axial location is known either a priori or from a conventional reconstruction method. There have been no applications of compressive holography for DIH-PTV of flows containing thousands of tracer particles. The primary barrier preventing such application is the high computational cost. For illustration, we consider a case with 1000 particles per image and 100 images each sized \(1024\times1024\) pixels with 1024 reconstruction planes (\(10^9\) voxels). The best reported speeds for parametric methods is approximately 4 seconds per particle (4.6 days for our example) \cite{Verrier2016}. A previous GPU-accelerated compressive holography implementation can reconstruct a volume up to \(1024\times1024\times10\) voxels in 7.6 seconds (22 hours for our example) \cite{Endo2016}. Other methods have even longer extrapolated times including 1000 days \cite{Jolivet2018} and 400 days (on modern hardware) \cite{Brady2009}. In addition to the time required for processing, memory requirements for CH place severe restrictions on the size of hologram that can be processed. Storage of the hypothetical test case hologram would require approximately 8 GB to store in memory (complex floating-point values, 8 bytes each). Several additional variables of this size are needed for CH algorithms. However, contemporary GPU memory is limited to approximately 12 GB for consumer hardware and memory transfers from the GPU to RAM are slow. Therefore, current application of CH must either limit the volume size to take advantage of the speed increase of GPU-acceleration or rely on the much larger RAM available on most desktop computers and rely on much slower CPU processing.

In the present study, we first summarize the fundamentals of CH. We then introduce our proposed method using fused lasso regularization and a sparse storage structure to enable processing of very large images in a realistic time (55 hours for the 100 image sequence of large holograms described above). We then provide several synthetic and experimental evaluation cases to demonstrate the quality and performance of the proposed method.

\section{Methodology}

We formulate the 3D reconstruction of the object volume as an inverse optimization problem, following the method of Brady, Endo, and others \cite{Brady2009, Endo2016}. The optimization problem formulation (\ref{eqn:opt_prob_formulation}) seeks to find the object field (\(x\)) that minimizes the difference between the observed hologram (\(b\)) and the estimated hologram produced from the object (\(\hat{b}=Hx\)). To ensure that the solution converges, we use a linearized form of the forward model for hologram formation which implicitly treats any nonlinear terms (i.e. twin image and cross interference) as noise.
\begin{equation}
\label{eqn:opt_prob_formulation}
    \hat{x} = \argmin_x{\left\{\norm{Hx-b}_2^2 + \lambda R(x) \equiv f(x) + g(x) \right\}}
\end{equation}
To avoid trivial solutions, a constraint must be applied to ensure a physically realistic solution. This constraint is implemented as the additional regularization term in (\ref{eqn:opt_prob_formulation}), \(\lambda R(x)=g(x)\). The form of this regularization function determines which properties of the solution will be enforced. The \(\ell^1\) norm (\ref{eqn:L1_norm}) enforces a sparse solution (i.e. one with few non-zero elements). This sparsity-based regularization has been demonstrated for holography by Denis et al. \cite{Denis2009} and Endo et al. \cite{Endo2016} who showcase its utility when the fraction of the sample volume occupied by objects is very small.
\begin{equation}
    \label{eqn:L1_norm}
    R(x)=\norm{x}_1 = \sum_{i=1}^N \abs{x_i}
\end{equation}

The Total Variation (TV) norm (\ref{eqn:TV_norm}) is the sum of the 1st order gradients over the image (size \(N_x\times N_y\)). It is naturally extensible to higher dimensions. TV regularization enforces a smooth solution (small gradients).
\begin{equation}
    \label{eqn:TV_norm}
    R(x)=\norm{x}_{TV} = \sum_{i=1}^{N_x}\sum_{j=1}^{N_y}\sqrt{(x_{i,j}-x_{i-1,j})^2 + (x_{i,j}-x_{i,j-1})^2}
\end{equation}

The TV approach has been used by Brady et al. \cite{Brady2009} and Endo et al. \cite{Endo2016} who demonstrate that it is superior to the \(\ell^1\) regularization for sufficiently large objects. However, we will see that TV regularization is substantially more computationally demanding that the \(\ell^1\) method.
We propose using the Fused Lasso (FL) regularization method (\ref{eqn:FL_norm}) which is a combination of the TV and \(\ell^1\) norms ("fusion" and "lasso" being alternative terms for TV and \(\ell^1\) respectively) \cite{Tibshirani2005}.
\begin{equation}
    \label{eqn:FL_norm}
    g(x) = \lambda R(x) = \lambda_{\ell^1}\norm{x}_1 + \lambda_{TV}\norm{x}_{TV}
\end{equation}
Solutions to the FL problem are both smooth and sparse while having some characteristics that make it less computationally demanding than TV. 

We solve the inverse problem using FISTA (Fast Iterative Shrinkage-Thresholding Algorithm) \cite{Beck2009a} as implemented for FASTA \cite{Goldstein2014}. This method is selected due to its high convergence rate, relative simplicity, and similarity to the approaches of Brady et al. \cite{Brady2009} and Endo et al. \cite{Endo2016}. FISTA is a proximal gradient method which makes use of the proximal operator \cite{Parikh2014} which can be interpreted as a gradient step with step size \(L\) (\ref{eqn:prox_definition}).
\begin{equation}
    \label{eqn:prox_definition}
    prox_{L}(v) = \argmin_x\left(g(x)+\frac{1}{2L}\norm{x-v}_2^2\right)\approx v-L\nabla g(x)
\end{equation}
FISTA has two steps: a shrinkage step using the proximal operator and an accelerated update using the previous estimate. The reader is referred to the references for further details on FISTA.
The shrinkage step of FISTA is by far the most computationally complex (the accelerated update uses only basic arithmetic), taking the form:
\begin{equation}
    x_k = prox_{\lambda L}\left(x_{k-1}-L\nabla f(x_{k-1})\right)
\end{equation}
As such, the computational cost of FISTA is closely linked to that of evaluating the proximal operator. For the \(\ell^1\) regularizer, the proximal operator has a simple closed-form solution as soft thresholding
\begin{equation}
    prox_{\lambda L}(v)=\left(1-\frac{\lambda L}{\abs{v}}\right)_+ sign(v)
\end{equation}
However, the proximal for the TV function does not have a closed-form solution and requires an iterative solution, for which we use the gradient projection method of Beck \& Teboulle \cite{Beck2009}. This method requires storage of each directional derivative for the duration of the iterations which may require a substantial amount of memory. 
The FL regularization function has the useful property that it is separable and can be computed by soft thresholding the solution to the TV problem (i.e. with \(\lambda_{\ell^1}=0\)). Because the non-sparse TV solution must first be computed before soft thresholding to produce the sparse FL solution, high memory requirements of the TV proximal still apply within each FISTA step even though the result is sparse. We limit our TV regularization to 2D planes which can be computed independently, reducing the memory requirements to those of a single plane. It is worth noting that prior compressive holography methods using TV regularization have reported only the 2D variant.

Because FISTA is an iterative solution method, the computational time required to process a single image may be relatively high. PTV requires processing thousands of large, well-resolved volumes. Therefore, it is crucial to reduce the processing time to a manageable level to enable application to real flow studies. We utilize a CUDA/C++ GPU implementation of our algorithm to accelerate the processing. Key components such as the fast Fourier transform (FFT) already have efficient GPU library implementations while the reconstruction kernel and proximal operator largely use highly parallelizable pixel-wise operations. However, a reasonably sized reconstruction volume (\(1024\times1024\times1024\) voxels) would require approximately 8 GB to store in memory (complex floating-point values, 8 bytes each). Several additional variables of this size are needed for FISTA. However, contemporary GPU memory is limited to approximately 12 GB which would place limits on the type of holograms which could be processed. In order to circumvent this challenge, we exploit the sparsity of the \(\ell^1\) and FL regularized solutions to dramatically reduce the memory requirements. We use the coordinate (commonly, COO) sparse matrix format to store all volume data during iterations. The COO format stores the indices (row and column) and value for each non-zero element in a plane. Because data is accessed per plane for both the forward and reverse propagation as well as the 2D TV proximal operator, each plane is independently indexed. The total storage for each non-zero element is thus 24 bytes (8 bytes per index, 8 bytes for complex floating-point value) compared to 8 bytes per element for a non-sparse structure. Thus, memory usage should be reduced as long as the data sparsity (number of zero elements divided by total) exceeds 67\%. Experience suggests that most PTV holograms have sparsity exceeding 99\% \cite{Toloui2015, Malek2004}.

The primary advantage of the compressive holography approach is that it produces very high SNR reconstructions that are more easily segmented for particle localization. In one sense, the sparsity regularization inherently separates objects (non-zero voxels) from the background (zero voxels), thus negating the need for complex volume normalization and SNR enhancement such as that used by Toloui et al. \cite{Toloui2017}. While these directly thresholded results are reasonable, we have found that two additional filters greatly reduce the instances of over-segmentation. The first is a very low intensity threshold on the order of \(1/256^{th}\) of the maximum intensity of the image. This value is selected as any values below it would be indistinguishable from zero when using a min-max scaling and 8 bit discretization for visualization. The second filter is a minimum object volume. This must be adjusted slightly depending on the size of the particles, noise level, and apparent elongation length. At this time, it is not directly linked to the true particle volume. Usually, objects of 5 voxels or fewer are treated as noise. While crucial for counting the number of particles in a single hologram, these parameters have minimal effect when applied to a sequence of images for which the particles are tracked because over-segmentation noise rarely persists for multiple frames. 

While compressive holography and inverse methods have existed for over a decade, this is the first application to 3D PTV. Previous uses of a parametric inverse method for particle tracking (\cite{Chareyron2012, Marie2017, Seifi2013a, Verrier2015, Verrier2016}) have tracked fewer than 10 particles concurrently. Furthermore, CH is usually used with a small number (\(\sim10\)) of reconstruction planes with a large spacing (\(\sim1\) mm) between planes. Here we demonstrate the ability to reconstruct volumes with over 1000 planes with cubic reconstructed voxels. The largest volume reconstructed by Endo et al. \cite{Endo2016} contained \(10^7\) voxels while our sparse representation enables reconstruction of volumes containing more than \(10^9\) voxels on a desktop computer. The use of the fused lasso regularization to enforce both smoothness and sparsity has not been previously demonstrated for compressive holography. To emphasize these distinctions, we refer to our method as a Regularized Inverse Holographic Volume Reconstruction (RIHVR, pronounced "river"). RIHVR dramatically increases the SNR of the reconstructed volume. This enables processing of high noise and high particle concentration holograms (both traits are common in DIH-PTV applications) that could not be reliably processed using existing methods. Because RIHVR does not assume a size or shape of the object, it can be used when the imaged particles are polydisperse or non-spherical. We next present several practical examples to demonstrate these capabilities.

\section{Demonstration Cases} \label{sec:demonstration}

To demonstrate that the proposed method is applicable to a variety of DIH-PTV cases, we present the results for processing four classes of holograms: an isolated nanowire, simulated tracer particles in isotropic turbulence, swimming microorganisms, and an experimental T-junction flow seeded with microfibers. The first case, the isolated nanowire, demonstrates improved 3D reconstruction of a continuous object with a significant 3D shape. Simulated holograms then provide a realistic flow case for which ground truth exists for the particle locations. The RIHVR method is evaluated against deconvolution (with inverse iterative particle extraction where applicable) which has been previously validated against conventional PIV and shown to provide substantial improvement over other DIH-PTV approaches \cite{Toloui2017}. A simple reconstruction method following the approach of Pan \& Meng \cite{Pan2003} (global thresholding followed by peak intensity depth localization) is also shown for comparison. Experimental holograms of swimming microorganisms and microfibers in a T-junction flow represent real measurement domains for which some flow behaviors are known from prior studies. These later cases demonstrate that RIHVR can be applied to broad measurement domains where other DIH-PTV methods fail.

\subsection{Isolated Nanowire}

\begin{figure}[h!]
\centering
\includegraphics[width=\textwidth]{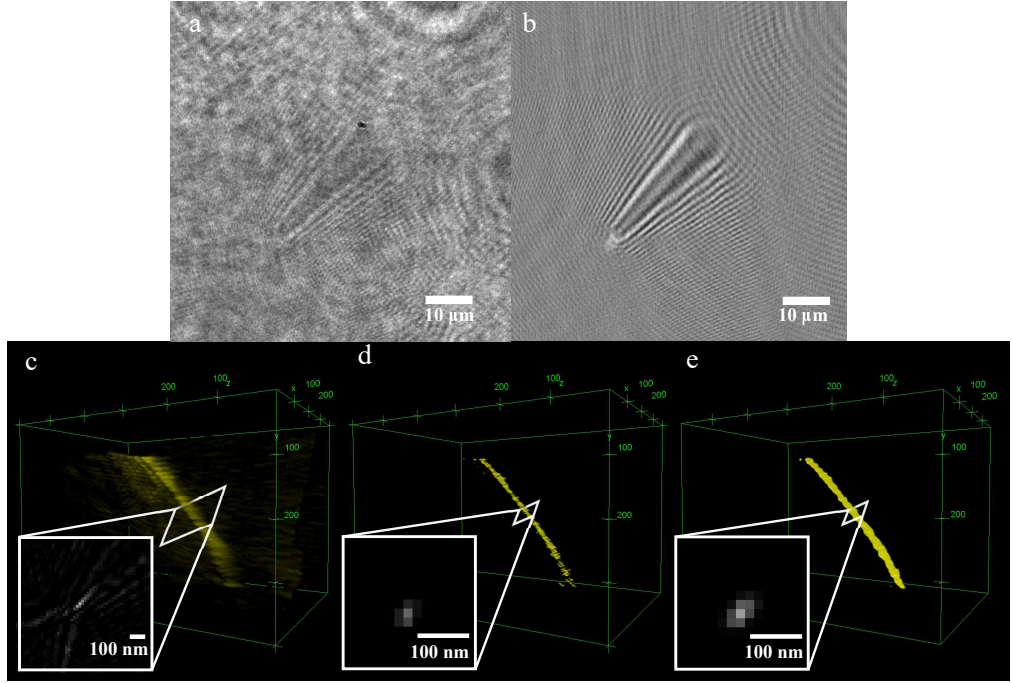}
\caption{(a) Recorded holographic image of a 90 nm Ag nanowire. (b) Hologram after image enhancement. (c) Volumetric reconstruction of the sample using the deconvolution method. (d) Reconstruction using RIHVR with sparsity (\(\ell^1\)) regularization. (e) Reconstruction using RIHVR with fused lasso (FL) regularization. For visualization, (c) uses the intensity as the transparency alpha value while (d) and (e) show all non-zero values at equal intensity.}
\label{fig:nanowire}
\end{figure}

A qualitative evaluation of the proposed inverse reconstruction method uses a silver nanowire in suspension. This is an example of a continuous object with significant extent in all three reconstruction dimensions. The length of the wire is not known \emph{a priori}. As such, a parametric inverse model such as the one used by Soulez et al. \cite{Soulez2007} is unsuitable. The sample is a suspension of 90 nm diameter Ag nanowires in isopropyl alcohol.
The illumination source is a 450 nm fiber-coupled laser diode (QPhotonics QFLD-450-10S), collimated using a Nikon CFI Plan Fluor 10X objective lens.
A Nikon CFI Apo TIRF 100X oil immersion microscopic objective and video camera (Andor Zyla 5.5 sCMOS) are used to image the sample. The recorded pixel size is 70 nm. The recorded image (\(2560\times2160\) pixels) is cropped to a \(1024\times1024\) pixel region around a selected nanowire to ensure that only a single object is in the image and to reduce unnecessary computational cost. Reconstruction is performed at 70 nm intervals (equal to the lateral pixel pitch) for a depth of 42 \(\mu\)m (600 planes). Measurement of similar samples using DIH has been undertaken by Dixon et al. \cite{Dixon2011} who measured the diffusion of nanowires and Kempkes et al. \cite{Kempkes2009} who demonstrated a \(2^\circ\) accuracy for the orientation of microfibers.
Unlike the prior methods, our approach does not assume a linear fiber and is suitable for measuring non-rigid wires.

The raw and enhanced holograms are shown in Figures \ref{fig:nanowire}a and \ref{fig:nanowire}b respectively alongside renderings of the reconstructed volumes produced using deconvolution, RIHVR with sparsity regularization, and RIHVR with fused lasso regularization. The figure shows that both regularization methods substantially reduce the DOF of the reconstruction.
Measured as the width at half the measured intensity averaged along the wire length, the DOF decreases from 1.97 \(\mu\)m using deconvolution to 0.63 \(\mu\)m and 0.89 \(\mu\)m using the sparsity and fused lasso regularization methods respectively. Similarly, a 99\% decrease in the segmented volume and 90\% decrease in the segmented cross-sectional area is observed between deconvolution and RIHVR (with similar reduction for both RIHVR regularizers). 

When comparing the results generated using the sparsity and fused lasso regularization methods, the smoothing effect of the fused lasso is apparent (Figure \ref{fig:nanowire}d and e). The fused lasso regularized results show fewer gaps in the wire profile and an overall more contiguous object. However, this comes at the cost of some expansion of the object and a slightly larger DOF. Interpolated cross-sections normal to the wire axis (insets in Figure \ref{fig:nanowire}) illustrate that both RIHVR approaches approximate the true circular shape of the wire.  Conversely, deconvolution (Figure \ref{fig:nanowire}c) produces an X-shaped cross-section characteristic of simple holographic reconstructions. RIHVR also demonstrates robustness to image noise. The raw image (Figure \ref{fig:nanowire}a) has a substantial amount of background noise and even enhancement via background removal does not produce a noise free image (Figure \ref{fig:nanowire}b). Additional fringe patterns -- caused by vibrations, fluctuations in illumination intensity, and out-of-view objects -- are visible in the enhanced image but do not result in artifacts in the reconstructed volume when using RIHVR.

\subsection{Synthetic Turbulent Flow}

Turbulent flows represent the most challenging case for 3D flow measurements as they are highly three-dimensional and involve velocity fluctuations across a broad dynamic range of scales. Here we assess the accuracy and limitations of our method using simulated holograms of a homogeneous isotropic turbulent flow. The simulated tracer particle trajectories are determined by querying the forced isotopic turbulence data from the Johns Hopkins Turbulence Database with Lagrangian particle tracking \cite{Li2008, Perlman2007, Yu2012}. 
The simulation domain is scaled to \(5\times5\times5\) mm\(^3\) and sampled with a nondimensional time step of 0.012 (60 DNS time steps) to capture 100 instants (image frames). The Reynolds number based on the domain size is 23,000 and the Kolmogorov length and time scales (smallest scales of turbulent fluctuations) are 67 \(\mu\)m and 3.7 frames respectively. The RMS velocity is 6.7 \(\mu\)m/frame. Maintaining the Reynolds number and using a low viscosity fluid (\(\nu=10^{-7}\) m\(^2\)/s), this corresponds to a frame rate of 75 kHz which is achievable with modern cameras.
The particles are initially randomly spatially distributed throughout the 3D domain and their positions at subsequent time steps are determined using a Lagrangian tracking method \cite{Yu2012}. A periodic boundary condition is applied to the particles to ensure that the number of objects in the field of view is constant (this is ignored during processing). The simulated holograms are \(512\times512\) pixel images with a 10 \(\mu\)m pixel size and 632 nm illumination wavelength. 
\begin{figure}[h!]
\centering
\includegraphics[width=\linewidth]{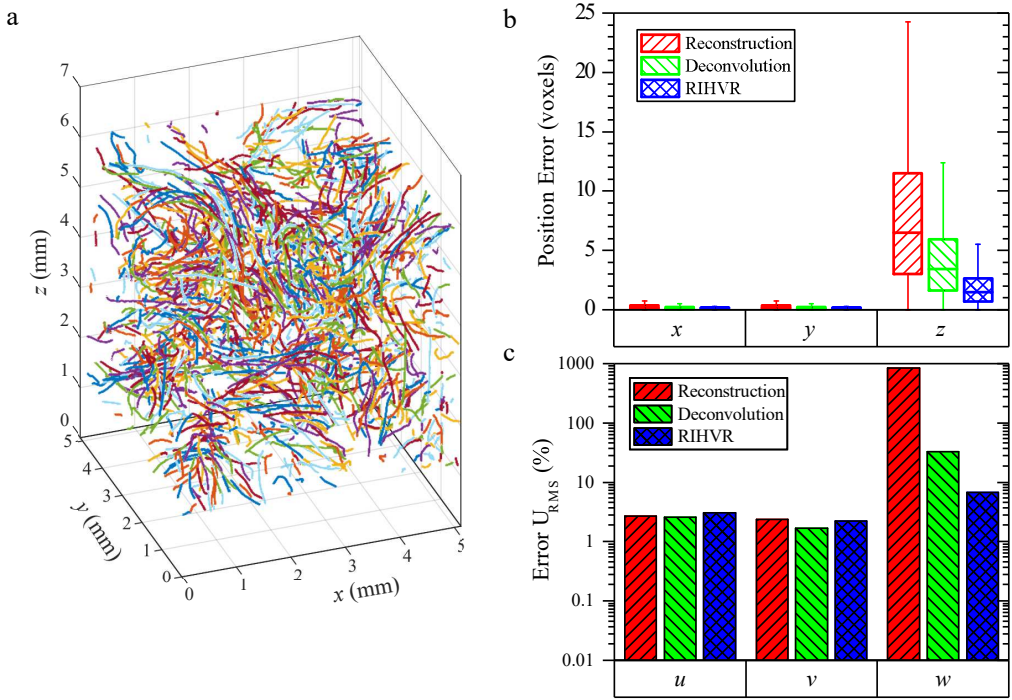}
\caption{(a) Smoothed 3D particle trajectories extracted from a synthetic hologram using RIHVR. (b) Localization error of tracked particles relative to their true locations. (c) Error in the RMS velocity components of the three test methods compared to ground truth.}
\label{fig:synthetic_accuracy}
\end{figure}

The reconstruction plane spacing is equal to the lateral pixel spacing (10 \(\mu\)m). The total size of the reconstructed volume is \(512\times512\times700\) voxels (\(1.8\times 10^8\)). 100 FISTA iterations are used with regularization parameters \(\lambda_{\ell^1}=0.5\) and \(\lambda_{FL}=0.2\). For segmentation, the minimum intensity threshold is set to \(2/256^{th}\) the maximum and the minimum volume is 5 voxels. The particle positions are estimated using the weighted centroid of each connected component and the particles are tracked using the method of Crocker \& Grier \cite{Crocker1996} with a maximum per frame displacement of 70 \(\mu\)m and a minimum tracked duration of 10 frames. Two alternative hologram reconstruction methods are presented for comparison. The first is a simple reconstruction method following the approach of Pan \& Meng \cite{Pan2003}. The second is the deconvolution method of Toloui \& Hong \cite{Toloui2015} with two passes of the inverse iterative particle extraction step. The particle trajectories for all methods are smoothed with a total variation filter. The tracked results using RIHVR are shown in Figure \ref{fig:synthetic_accuracy}a.

To evaluate the localization error, extracted particles are matched to their true location using a nearest neighbor method \cite{Crocker1996}. The resulting error distributions in each dimension are summarized in Figure \ref{fig:synthetic_accuracy}b. For all three methods, the error in \(x\) and \(y\) is very small (smaller than the pixel size). However, the error in \(z\) is substantially greater, demonstrating the DOF problem. Comparing the three methods, the \(75^{th}\) percentile decreases from 11.5 voxels using reconstruction to 6 voxels using deconvolution and 3.5 voxels with RIHVR. The same trends are seen at the other percentiles as well. Thus, RIHVR produces a 40\% improvement in longitudinal localization over the prior best method and a 70\% improvement over simple reconstruction.
\begin{figure}[h!]
\centering
\includegraphics[width=\linewidth]{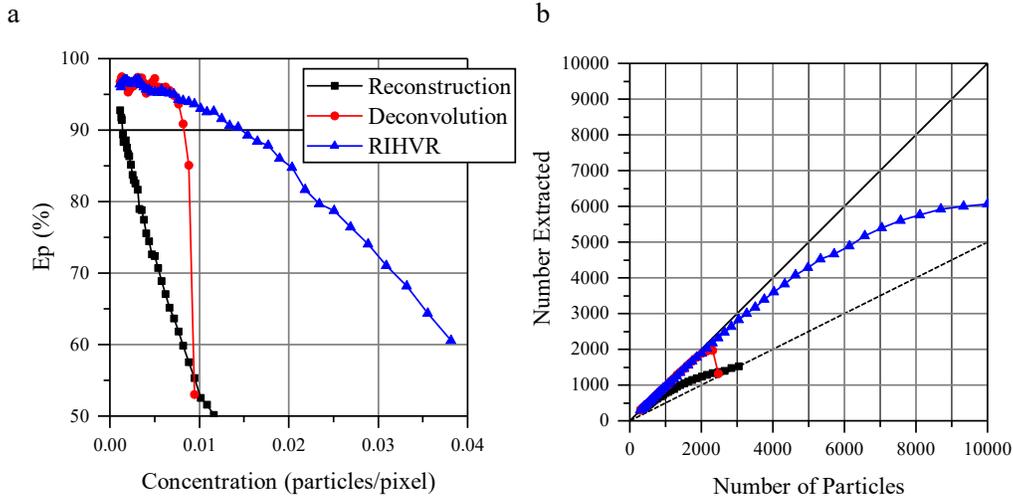}
\caption{(a) Extraction rate for each method for increasing particle concentrations. (b) The number of particles which can be accurately extracted is higher for RIHVR than the other methods. Solid line is 100\% EP, dashed line is 50\%.}
\label{fig:synthetic_concentration}
\end{figure}

For turbulence measurements, it is common to measure Reynolds stresses which are velocity fluctuation statistics \cite{Sheng2009}. Here, we present measurements of the root-mean-square (RMS) velocity (Figure \ref{fig:synthetic_accuracy}c). This is comparable to Reynolds stress when the mean is zero (as it is for this flow) while maintaining intuitive meaning for applications other than flow measurement. For this flow in the period during which particles are simulated, the true RMS velocities averaged over the whole volume are (6.3, 7.6, 6.2) voxels/frame in the \(u\), \(v\), and \(w\) directions respectively. The trajectory smoothing produces a 3\% error in the \(u_{RMS}\) and \(v_{RMS}\) measurements but significantly reduces spurious fluctuations in \(z\). Using reconstruction, the measured \(w_{RMS}\) differs from the true value by over 800\%. This is reduced to 30\% using deconvolution. However, this is still unacceptably high for real measurements. The error using RIHVR is only 7\% which is substantially better and only 2\(\times\) greater than the error in the \(u\) and \(v\) measurements.

As the velocity vector spacing in PTV is directly related to the particle spacing, the maximum particle concentration is a critical concern for many PTV measurements. The quality of recorded holograms depends on several factors including the particle concentration, size, volume depth, and image resolution. In general, the extraction rate (\(Ep\), number of correctly extracted particles divided by true particle count) decreases with increasing concentration. Here we use the number of particles per pixel to scale the concentration because it allows for the most direct comparison with the literature. It has previously been shown that the commonly used shadow density \ref{eqn:shadow_density} does not completely explain the extraction rate in all situations \cite{Malek2004, Toloui2015}. For comparison, Toloui et al. \cite{Toloui2017} performed measurements with a concentration of 0.0035 particles/pixel while other sources used significantly lower concentrations. Using RIHVR, concentrations of up to 0.035 particles/pixel can be processed while maintaining \(Ep>60\%\) (Figure \ref{fig:synthetic_concentration}a). The increased number of extractable particles enabled by RIHVR (Figure \ref{fig:synthetic_concentration}b) enables increased resolution in velocimetry applications and higher particle concentrations in other applications including studies of biological flows and fluid-particle interaction where high concentration may be crucial for the sample being studied. An example of such a case is given in section \ref{sec:swimming_algae}.

\subsection{Swimming Algae} \label{sec:swimming_algae}

One practical application of DIH-PTV is the study of microorganism swimming behaviors. Previous studies have used small sample volumes (~\(\sim0.05 \mu\)L) in order to measure the large cell concentration present in cultures (\(\sim10^6\) cells/mL) \cite{Bedrossian2018,Molaei2014}. Here we demonstrate that RIHVR is superior to prior DIH algorithms for these experiments. We also demonstrate the ability to record and process much larger sample volumes (\(\sim1 \mu\)L) which could enable new scientific studies.

\begin{figure}[h!]
\centering
\includegraphics[width=\linewidth]{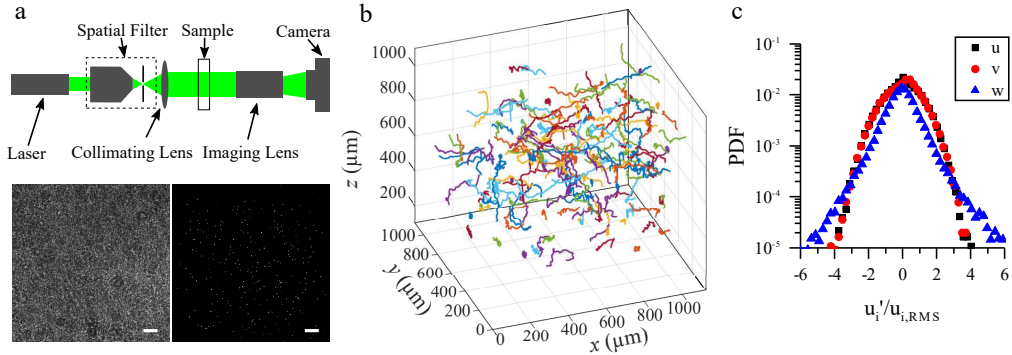}
\caption{(a) DIH imaging system, recorded hologram, XY projection of processed reconstruction. Scale bars are 100 \(\mu\)m. (b) Subset (25\%) of tracked particles. (c) PDF of velocity fluctuations in each direction. The w distribution has longer tails and a sharper peak but is not substantially wider than the other two components.}
\label{fig:dunaliella}
\end{figure}

The alga \emph{Dunaliella primolecta} is a unicellular species which can be used for biofuel production \cite{Amaro2011}. Cells have a length of 10 \(\mu\)m and swim using two flagella \cite{Chengala2013a}. In this study, \emph{D. primolecta} is grown at \(37^\circ\)C in a growth medium. Manual concentration measurements using a microscope indicate that the sample has a concentration of \(1.8\times10^6\) cells/mL. The sample container is a \(10\times30\times1\) mm\(^3\) glass cuvette. Holograms are recorded at 100 Hz using a \(2048\times1088\) pixel sensor (Flare 2M360-CL). The sensor pixel size is 5 \(\mu\)m and a 5x microscopic objective is used. The recorded sample volume is \(2.05\times1.09\times1\) mm\(^3\). For simplicity and speed, the recorded image is cropped to a size of \(1024\times1024\) pixels (\(1\times1\) mm\(^2\)). The light source is a 532 nm diode laser (Thorlabs CPS532) which is expanded and filtered with a spatial filter (see Figure \ref{fig:dunaliella}a). 
While the number of particles per pixel is relatively low for this sample (0.002), the particles are large enough that the shadow density (\ref{eqn:shadow_density}) becomes significant, \(s_d=18\%\). The maximum shadow density used by Toloui et al. \cite{Toloui2017} was 10.5\% using deconvolution while Malek et al. \cite{Malek2004} achieved an extraction rate of only 20\% for \(s_d=10\%\). Reducing the measurement depth can enable holograms to be processed using conventional methods \cite{Chengala2013a}, but risks introducing wall effects that influence the behavior. Similarly, we have found that dilution of the sample changes the cell swimming behavior. Therefore, high concentration holograms -- which can only be processed using RIHVR -- are important to these microbiological studies. Studies of microorganism behavior using non-holographic methods are challenging because their 3D motion leads to low residence time in a microscopic depth of field and size constraints make multi-camera imaging difficult.

The holographic volume is reconstructed with 270 planes, separated by 3 \(\mu\)m, with the volume confirmed to include both walls of the cuvette. The regularization parameters are \(\lambda_{\ell^1}=0.1\) and \(\lambda_{TV}=0.1\) with 100 FISTA iterations and 5 gradient projection steps to evaluate the TV proximal operator. A sequence of 2000 frames is analyzed. Images are preprocessed by removing the mean of the sequence using the method of \cite{Seifi2013} (subtraction followed by division by the square root). A sliding window of 151 images (1.5 sec) is used to compute the mean background in order to reduce the effect of cells starting or stopping their motion.

\begin{figure}[h!]
\centering
\includegraphics{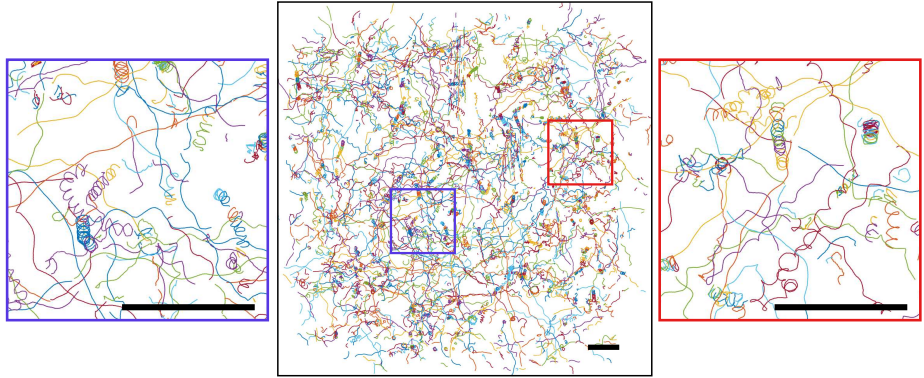}
\caption{2D view of the reconstructed cell trajectories showing complex behaviors Also illustrates the true cell concentration of processed volume. Scale bars are 100 \(\mu\)m.}
\label{fig:dunaliella_zoomed}
\end{figure}

RIHVR detects and tracks an average of 294 objects per frame. This is dramatically lower than the expected count of 2000 cells/frame from the concentration measurement. However, a substantial number of particles are seen to remain stationary on the two walls. These are treated as background noise and are removed during the image enhancement. A selection of 3D tracks is shown in Figure \ref{fig:dunaliella}b. For clarity, only a subset of 25\% of the tracked data is shown in Figure \ref{fig:dunaliella} while the full density is shown in Figure \ref{fig:dunaliella_zoomed}. The cell trajectories have been smoothed using a Savitzky-Golay filter of 20 frames. The frame rate is sufficiently high that this filter does not suppress any real motions. Under the assumption that cell swimming motions are isotropic, the probability distribution functions (PDF) of velocity fluctuations (normlized by the RMS velocity) are expected to coincide for each component. Figure \ref{fig:dunaliella}c shows that while the \(u\) and \(v\) velocities are in good agreement, the same is not true for \(w\), even after smoothing. This indicates that the DOF problem is not entirely eliminated for this extremely noisy case. However, gross motions in the longitudinal direction are visible and fine scale complex behaviors such has helical swimming can be seen from an enlarged view of the sample (Figure \ref{fig:dunaliella_zoomed}).

\subsection{Rotating Rods in Flow}

In addition to improvements to the measurement accuracy and seeding density limits, RIHVR enables the measurement of complex shapes as previously illustrated in Figure \ref{fig:nanowire}. Here we present a flow case where the seeding particles are rods rather than the usual spherical tracers. Using RIHVR, we are able to extract both the location and orientation of each rod and track their evolution in the flow. This type of multimodal measurement using a single camera has not been previously reported.

To demonstrate this measurement, we use the T junction flow of the type studied by \cite{Vigolo2014} which occurs frequently in industrial and biological flows. The rotation and alignment of fibers in flow have been extensively studied for target applications including paper manufacturing and microorganism alignment (see for example \cite{Shin2005, Katz2006, Parsa2012}). The fibers used for the present study are marketed as an additive to strengthen composite materials, where the alignment of the fibers may have an impact on the material properties. Prior experimental work has either been restricted to 2D measurements \cite{Parsa2011, Katz2006} or multi-camera 3D measurements of individual fibers \cite{Parsa2012}. Holography is a valuable alternative when the motion is three-dimensional, seeding density is high, or optical access is restricted. 

The experimental channel has a square cross-section with a side length of 1 mm. The junction is at a right angle and all three branches (inlet and 2 outlets) have the same geometry. The inlet flow rate is 1000 mL/hr which corresponds to a Reynolds number \(Re=290\). The seeding particles are 7 \(\mu\)m diameter SiC (\(\rho=3\) g/cm\(^3\)) microfibers (Haydale Technologies) with an aspect ratio of 10.
The response time of the particle, computed using the equivalent diameter, is \(\tau_p=14\) \(\mu\)s. The characteristic time of the flow is \(\tau_f=1.9\) ms. The resultant Stokes number is \(St_k=0.007\) indicating that the particles will trace the flow.
A high speed video camera (NAC Memrecam HX-5) is used to record the holograms at 6000 Hz. A microscopic objective (Edmund Optics, 10x, NA=0.45) is used to image the sample, resulting in a \(1024\times1024\) image with a pixel size of 0.91 \(\mu\)m. The light source is a spatially filtered HeNe laser (\(\lambda=632\) nm).
The FL regularization method is used with \(\lambda_{\ell^1}=0.1\) and \(\lambda_{TV}=0.12\). 110 reconstruction planes are used with a spacing of 9.1 \(\mu\)m (\(10\times\) the pixel pitch). An intensity-weighted principle component analysis is used to determine the orientation of the fibers (similar to the method of \cite{Kempkes2009}).

\begin{figure}[h!]
\centering
\includegraphics{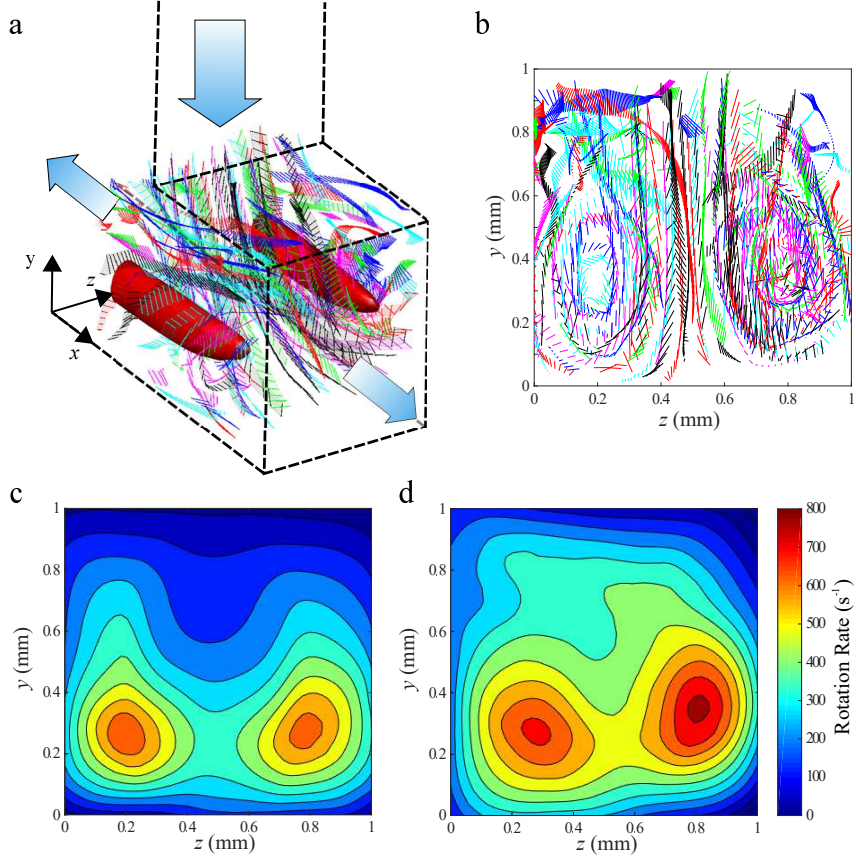}
\caption{(a) Visualization of the measured rod trajectories. Solid lines show the measured 3D orientation of the rods. Colors indicate individual particles. Vorticity isosurfaces (\(\omega=3000 s^{-1}\)) are from the CFD simulation. (b) View of the experimental fiber orientations in the yz plane. (c) Contour map of the particle rotation rate (\(s^{-1}\)) expected from the simulation. (d) Measured 3D particle rotation rate.}
\label{fig:tjunction}
\end{figure}

For validation of the flow field, the flow (absent any particles) is simulated using ANSYS Fluent (ANSYS, Inc.), with the results found to be in agreement with the simulations of \cite{Vigolo2014} and the experimental particle pathlines. The fiber rotation rate is modeled using the Jeffery equation in the limit where the particle aspect ratio is \(\gg\)1 \cite{Jeffery1922, Marcus2014}:
\begin{equation}
    \label{eqn:jeffery}
    \dot{p}_i=\boldsymbol{\Omega}_{ij}p_j + (\boldsymbol{S}_{ij}p_j-p_ip_j\boldsymbol{S}_{jk}p_k)
\end{equation}
Where \(\boldsymbol{p}\) is a unit vector aligned with the particle axis, \(\boldsymbol{\Omega}\) is the rotation tensor, and \(\boldsymbol{S}\) is the strain rate tensor. Because the particle rotation rate is coupled to the orientation, the rotation rates for the simulation (Figure \ref{fig:tjunction}c) assume that the particles are initially aligned with the inlet flow direction.

The experimental fiber orientations are shown in Figure \ref{fig:tjunction}a along with vorticity isosurfaces from the simulation which illustrate that the principle flow features predicted by the simulation (two vortices aligned with \(x\)) are present in the experiment. A 2D projection of the fibers is shown in Figure \ref{fig:tjunction}b. The optical reconstruction direction corresponds to the crossflow (\(z\)) axis in these figures. The clear appearance of the two counter-rotating vortices demonstrates that RIHVR sufficiently reduces the DOF to enable the recovery of this 3D flow feature. Additionally, changes in the orientation of the fibers can be seen. The measured rotation rate (\(\abs{\dot{p}}\)) is higher near the centers of the vortices (Figure \ref{fig:tjunction}d) which matches the expected behavior from the simulation (Figure \ref{fig:tjunction}c). The peak rotations rates are accurate to 30\% while the location of the peaks are accurate within 0.1 mm.
Some discrepancies between the measured and predicted rotation rates can be attributed to misalignment between the two domains. 
Because the rotation rate is dependent on the velocity gradients (which have substantial variation) and on the particle location and orientation (which have some measurement uncertainty), even small misalignment of the two domains would cause deviations between the two results. 
Non-ideal flow conditions, including unsteadiness and geometrical imperfections, could cause differences in the locations of the vortices and the peak rotation magnitude. The accuracy of the machining process used to make the channel is \(\pm 0.05\) mm which is comparable to the peak location error. Since the flow rate is constant, uncertainty in the channel geometry also produces uncertainty in the inlet flow velocity. Finally, the Jeffery equation (\ref{eqn:jeffery}) used to predict the particle rotation rate assumes non-inertial ellipsoidal particles in Stokes flow. The true particle motion is expected to deviate slightly from this idealization.
Given these uncertainties, we conclude that the agreement between the simulation and experimental results is adequate for demonstrating that RIHVR enables direct 3D particle rotation rate measurements.

\section{Conclusions}

We have demonstrated the application of CH techniques to volumetric reconstruction, using the presented RIHVR approach for the reconstruction and tracking of 3D particle fields. The reconstructed volumes are both sparse and smooth, assumptions that apply equally for most particle tracking applications. The use of GPU processing and sparse storage enable the reconstruction of volumes containing over \(10^9\) voxels which is orders of magnitude larger than previously reported for any CH method. RIHVR provides a substantial improvement in the longitudinal position and displacement measurement accuracy in addition to an increase in the particle concentration limit. These improved capabilities have allowed the extension of DIH-PTV to the tracking of a dense culture of microorganisms and measuring the orientation of microfibers in 3D flow. RIHVR is a broadly applicable approach capable of enabling low-cost 3D measurements for wide-ranging applications.

\section*{Funding}
National Science Foundation (NSF) (1427014); University of Minnesota Informatics Institute.

\section*{Acknowledgments}
We acknowledge the assistance of Santosh Kumar and Jiaqi You in performing measurements presented here.


\begin{thebibliography}{10}
\newcommand{\enquote}[1]{``#1''}

\bibitem{Poon2014}
T.-C. Poon and J.-P. Liu, \emph{{Introduction to Modern Digital Holography with
  MATLAB}} (Cambridge University Press, 2014).

\bibitem{Katz2010}
J.~Katz and J.~Sheng, \enquote{{Applications of Holography in Fluid Mechanics
  and Particle Dynamics},} {\protect\JournalTitle{Annual Review of Fluid
  Mechanics}} \textbf{42}, 531--555 (2010).

\bibitem{Yu2014}
X.~Yu, J.~Hong, C.~Liu, and M.~K. Kim, \enquote{{Review of digital holographic
  microscopy for three-dimensional profiling and tracking},}
  {\protect\JournalTitle{Optical Engineering}} \textbf{53}, 112306--112306
  (2014).

\bibitem{Seifi2013}
M.~Seifi, C.~Fournier, N.~Grosjean, L.~M{\'{e}}{\`{e}}s, J.-L. Mari{\'{e}}, and
  L.~Denis, \enquote{{Accurate 3D tracking and size measurement of evaporating
  droplets using in-line digital holography and "inverse problems"
  reconstruction approach},} {\protect\JournalTitle{Optics Express}}
  \textbf{21}, 27964 (2013).

\bibitem{Molaei2014}
M.~Molaei, M.~Barry, R.~Stocker, and J.~Sheng, \enquote{{Failed escape: Solid
  surfaces prevent tumbling of Escherichia coli},}
  {\protect\JournalTitle{Physical Review Letters}} \textbf{113}, 1--6 (2014).

\bibitem{Sheng2009}
J.~Sheng, E.~Malkiel, and J.~Katz, \enquote{{Buffer layer structures associated
  with extreme wall stress events in a smooth wall turbulent boundary layer},}
  {\protect\JournalTitle{Journal of Fluid Mechanics}} \textbf{633}, 17--60
  (2009).

\bibitem{Zhang2015}
C.~Zhang, R.~Miorini, and J.~Katz, \enquote{{Integrating Mach-Zehnder
  interferometry with TPIV to measure the time-resolved deformation of a
  compliant wall along with the 3D velocity field in a turbulent channel
  flow},} {\protect\JournalTitle{Experiments in Fluids}} \textbf{56}, 1--22
  (2015).

\bibitem{Toloui2019}
M.~Toloui, A.~Abraham, and J.~Hong, \enquote{{Experimental investigation of
  turbulent flow over surfaces of rigid and flexible roughness},}
  {\protect\JournalTitle{Experimental Thermal and Fluid Science}} \textbf{101},
  263--275 (2019).

\bibitem{Beals2015a}
M.~J. Beals, J.~P. Fugal, R.~A. Shaw, J.~Lu, S.~M. Spuler, and J.~L. Stith,
  \enquote{{Holographic measurements of inhomogeneous cloud mixing at the
  centimeter scale},} {\protect\JournalTitle{Science}} \textbf{350}, 87--90
  (2015).

\bibitem{Ozcan2016}
A.~Ozcan and E.~Mcleod, \enquote{{Lensless Imaging and Sensing},}
  {\protect\JournalTitle{Annu. Rev. Biomed. Eng}} \textbf{18}, 77--102 (2016).

\bibitem{Wu2017}
Y.-C. Wu, A.~Shiledar, Y.-C. Li, J.~Wong, S.~Feng, X.~Chen, C.~Chen, K.~Jin,
  S.~Janamian, Z.~Yang, Z.~S. Ballard, Z.~G{\"{o}}r{\"{o}}cs, A.~Feizi, and
  A.~Ozcan, \enquote{{Air quality monitoring using mobile microscopy and
  machine learning},} {\protect\JournalTitle{Light: Science {\&} Applications}}
  \textbf{6}, e17046 (2017).

\bibitem{Hinsch2002}
K.~Hinsch, \enquote{{Holographic particle image velocimetry},}
  {\protect\JournalTitle{Measurement Science and Technology}} \textbf{13},
  61--72 (2002).

\bibitem{Toloui2017}
M.~Toloui, K.~Mallery, and J.~Hong, \enquote{{Improvements on digital inline
  holographic PTV for 3D wall-bounded turbulent flow measurements},}
  {\protect\JournalTitle{Measurement Science and Technology}} \textbf{28}
  (2017).

\bibitem{Ling2016}
H.~Ling, S.~Srinivasan, K.~Golovin, G.~H. McKinley, A.~Tuteja, and J.~Katz,
  \enquote{{High-resolution velocity measurement in the inner part of turbulent
  boundary layers over super-hydrophobic surfaces},}
  {\protect\JournalTitle{Journal of Fluid Mechanics}} \textbf{801}, 670--703
  (2016).

\bibitem{Malek2004}
M.~Malek, D.~Allano, S.~Co{\"{e}}tmellec, and D.~Lebrun, \enquote{{Digital
  in-line holography: influence of the shadow density on particle field
  extraction.}} {\protect\JournalTitle{Optics express}} \textbf{12}, 2270--9
  (2004).

\bibitem{Chen2015}
W.~Chen, L.~Tian, S.~Rehman, Z.~Zhang, H.~P. Lee, and G.~Barbastathis,
  \enquote{{Empirical concentration bounds for compressive holographic bubble
  imaging based on a Mie scattering model},} {\protect\JournalTitle{Optics
  Express}} \textbf{23}, 4715 (2015).

\bibitem{Zhang1997}
J.~Zhang, B.~Tao, and J.~Katz, \enquote{{Turbulent flow measurement in a square
  duct with hybrid holographic PIV},} {\protect\JournalTitle{Experiments in
  Fluids}} \textbf{23}, 373--381 (1997).

\bibitem{Kebbel1999}
V.~Kebbel, M.~Adams, H.-J. Hartmann, and W.~J{\"{u}}ptner, \enquote{{Digital
  holography as a versatile optical diagnostic method for microgravity
  experiments},} {\protect\JournalTitle{Measurement Science and Technology}}
  \textbf{10}, 893--899 (1999).

\bibitem{Buchmann2013}
N.~A. Buchmann, C.~Atkinson, and J.~Soria, \enquote{{Ultra-high-speed
  tomographic digital holographic velocimetry in supersonic particle-laden jet
  flows},} {\protect\JournalTitle{Measurement Science and Technology}}
  \textbf{24} (2013).

\bibitem{Gao2018}
J.~Gao and J.~Katz, \enquote{{Self-calibrated microscopic dual-view tomographic
  holography for 3D flow measurements},} {\protect\JournalTitle{Opt. Express}}
  \textbf{26}, 16708--16725 (2018).

\bibitem{Cao2008}
L.~Cao, G.~Pan, J.~de~Jong, S.~Woodward, and H.~Meng, \enquote{{Hybrid digital
  holographic imaging system for three-dimensional dense particle field
  measurement.}} {\protect\JournalTitle{Applied optics}} \textbf{47},
  4501--4508 (2008).

\bibitem{Allano2013}
D.~Allano, M.~Malek, F.~Walle, F.~Corbin, G.~Godard, S.~Co{\"{e}}tmellec,
  B.~Lecordier, J.-m. Foucaut, and D.~Lebrun, \enquote{{Three-dimensional
  velocity near-wall measurements by digital in-line holography: calibration
  and results},} {\protect\JournalTitle{Applied Optics}} \textbf{52}, A9
  (2013).

\bibitem{Talapatra2013}
S.~Talapatra and J.~Katz, \enquote{{Three-dimensional velocity measurements in
  a roughness sublayer using microscopic digital in-line holography and optical
  index matching},} {\protect\JournalTitle{Measurement Science and Technology}}
  \textbf{24}, 024004 (2013).

\bibitem{Lindensmith2016}
C.~A. Lindensmith, S.~Rider, M.~Bedrossian, J.~K. Wallace, E.~Serabyn, G.~M.
  Showalter, J.~W. Deming, and J.~L. Nadeau, \enquote{{A Submersible, Off-Axis
  Holographic Microscope for Detection of Microbial Motility and Morphology in
  Aqueous and Icy Environments},} {\protect\JournalTitle{Plos One}}
  \textbf{11}, e0147700 (2016).

\bibitem{Kuhn2014}
J.~K{\"{u}}hn, B.~Niraula, K.~Liewer, J.~{Kent Wallace}, E.~Serabyn, E.~Graff,
  C.~Lindensmith, and J.~L. Nadeau, \enquote{{A Mach-Zender digital holographic
  microscope with sub-micrometer resolution for imaging and tracking of marine
  micro-organisms},} {\protect\JournalTitle{Review of Scientific Instruments}}
  \textbf{85} (2014).

\bibitem{Burns2007}
N.~Burns and J.~Watson, \enquote{{Data Extraction from Underwater Holograms of
  Marine Organisms},} {\protect\JournalTitle{OCEANS 2007 - Europe}} pp. 1--6
  (2007).

\bibitem{Wilson2012}
L.~Wilson and R.~Zhang, \enquote{{3D Localization of weak scatterers in digital
  holographic microscopy using Rayleigh-Sommerfeld back-propagation},}
  {\protect\JournalTitle{Optics Express}} \textbf{20}, 16735 (2012).

\bibitem{Denis2005}
L.~Denis, C.~Fournier, T.~Fournel, and C.~Ducottet, \enquote{{Twin-image noise
  reduction by phase retrieval in in-line digital holography},} in
  \emph{Proceedings of SPIE,}  M.~Papadakis, A.~F. Laine, and M.~A. Unser, eds.
  (2005), 2, p. 59140J.

\bibitem{Latychevskaia2007}
T.~Latychevskaia and H.-W. Fink, \enquote{{Solution to the Twin Image Problem
  in Holography},} {\protect\JournalTitle{Society}} \textbf{233901}, 1--4
  (2007).

\bibitem{Dixon2011}
L.~Dixon, F.~C. Cheong, and D.~G. Grier, \enquote{{Holographic deconvolution
  microscopy for high-resolution particle tracking},}
  {\protect\JournalTitle{Optics Express}} \textbf{19}, 16410 (2011).

\bibitem{Latychevskaia2010}
T.~Latychevskaia, F.~Gehri, and H.-W. Fink, \enquote{{Depth-resolved
  holographic reconstructions by three-dimensional deconvolution},}
  {\protect\JournalTitle{Optics Express}} \textbf{18}, 22527 (2010).

\bibitem{Latychevskaia2014}
T.~Latychevskaia and H.-W. Fink, \enquote{{Holographic time-resolved particle
  tracking by means of three-dimensional volumetric deconvolution},}
  {\protect\JournalTitle{Optics Express}} \textbf{22}, 20994 (2014).

\bibitem{Toloui2015}
M.~Toloui and J.~Hong, \enquote{{High fidelity digital inline holographic
  method for 3D flow measurements},} {\protect\JournalTitle{Optics Express}}
  \textbf{23}, 27159 (2015).

\bibitem{Soulez2007a}
F.~Soulez, L.~Denis, C.~Fournier, E.~Thi{\'{e}}baut, and C.~Goepfert,
  \enquote{{Inverse-problem approach for particle digital holography: accurate
  location based on local optimization.}} {\protect\JournalTitle{Journal of the
  Optical Society of America. A, Optics, image science, and vision}}
  \textbf{24}, 1164--1171 (2007).

\bibitem{Brady2009}
D.~J. Brady, K.~Choi, D.~L. Marks, R.~Horisaki, and S.~Lim,
  \enquote{{Compressive holography.}} {\protect\JournalTitle{Optics express}}
  \textbf{17}, 13040--13049 (2009).

\bibitem{Denis2009}
L.~Denis, D.~Lorenz, E.~Thi{\'{e}}baut, C.~Fournier, and D.~Trede,
  \enquote{{Inline hologram reconstruction with sparsity constraints.}}
  {\protect\JournalTitle{Optics letters}} \textbf{34}, 3475--3477 (2009).

\bibitem{Jolivet2018}
F.~Jolivet, F.~Momey, L.~Denis, L.~M{\'{e}}{\`{e}}s, N.~Faure, N.~Grosjean,
  F.~Pinston, J.-L. Mari{\'{e}}, and C.~Fournier, \enquote{{Regularized
  reconstruction of absorbing and phase objects from a single in-line hologram,
  application to fluid mechanics and micro-biology},}
  {\protect\JournalTitle{Optics Express}} \textbf{26}, 8923 (2018).

\bibitem{Verrier2016}
N.~Verrier, N.~Grosjean, E.~Dib, L.~M{\'{e}}{\`{e}}s, C.~Fournier, and J.-L.
  Mari{\'{e}}, \enquote{{Improvement of the size estimation of 3D tracked
  droplets using digital in-line holography with joint estimation
  reconstruction},} {\protect\JournalTitle{Measurement Science and Technology}}
  \textbf{27}, 045001 (2016).

\bibitem{Endo2016}
Y.~Endo, T.~Shimobaba, T.~Kakue, and T.~Ito, \enquote{{GPU-accelerated
  compressive holography},} {\protect\JournalTitle{Optics Express}}
  \textbf{24}, 8437 (2016).

\bibitem{Tibshirani2005}
R.~Tibshirani, M.~Saunders, S.~Rosset, J.~Zhu, and K.~Knight,
  \enquote{{Sparsity and Smoothness via the Fused Lasso},}
  {\protect\JournalTitle{Journal of the Royal Statiscical Society. Series B
  (Statistical Methodology)}} \textbf{67}, 91--108 (2005).

\bibitem{Beck2009a}
A.~Beck and M.~Teboulle, \enquote{{Fast gradient-based algorithms for
  constrained total variation image denoising and deblurring problems},}
  {\protect\JournalTitle{IEEE Transactions on Image Processing}} \textbf{18},
  2419--2434 (2009).

\bibitem{Goldstein2014}
T.~Goldstein, C.~Studer, and R.~Baraniuk, \enquote{{A Field Guide to
  Forward-Backward Splitting with a FASTA Implementation},}
  {\protect\JournalTitle{arXiv:1411.3406}} p.~25 (2014).

\bibitem{Parikh2014}
N.~Parikh and S.~Boyd, \enquote{{Proximal Algorithms},}
  {\protect\JournalTitle{Foundations and Trends in Optimization}} \textbf{1},
  127--239 (2014).

\bibitem{Beck2009}
A.~Beck and M.~Teboulle, \enquote{{A Fast Iterative Shrinkage-Thresholding
  Algorithm},} {\protect\JournalTitle{Society for Industrial and Applied
  Mathematics Journal on Imaging Sciences}} \textbf{2}, 183--202 (2009).

\bibitem{Chareyron2012}
D.~Chareyron, J.~L. Mari{\'{e}}, C.~Fournier, J.~Gire, N.~Grosjean, L.~Denis,
  M.~Lance, and L.~M{\'{e}}{\`{e}}s, \enquote{{Testing an in-line digital
  holography 'inverse method' for the Lagrangian tracking of evaporating
  droplets in homogeneous nearly isotropic turbulence},}
  {\protect\JournalTitle{New Journal of Physics}} \textbf{14} (2012).

\bibitem{Marie2017}
J.~L. Mari{\'{e}}, T.~Tronchin, N.~Grosjean, L.~M{\'{e}}{\`{e}}s, O.~C.
  {\"{O}}zt{\"{u}}rk, C.~Fournier, B.~Barbier, and M.~Lance, \enquote{{Digital
  holographic measurement of the Lagrangian evaporation rate of droplets
  dispersing in a homogeneous isotropic turbulence},}
  {\protect\JournalTitle{Experiments in Fluids}} \textbf{58}, 1--13 (2017).

\bibitem{Seifi2013a}
M.~Seifi, \enquote{{Signal processing methods for fast and accurate
  reconstruction of digital holograms},} Ph.D. thesis (2013).

\bibitem{Verrier2015}
N.~Verrier, C.~Fournier, and T.~Fournel, \enquote{{3D tracking the Brownian
  motion of colloidal particles using digital holographic microscopy and joint
  reconstruction},} {\protect\JournalTitle{Applied Optics}} \textbf{54}, 4996
  (2015).

\bibitem{Pan2003}
G.~Pan and H.~Meng, \enquote{{Digital holography of particle fields :
  reconstruction by use of complex amplitude},} {\protect\JournalTitle{Applied
  Optics}} \textbf{42}, 827--833 (2003).

\bibitem{Soulez2007}
F.~Soulez, L.~Denis, E.~Thi{\'{e}}baut, C.~Fournier, and C.~Goepfert,
  \enquote{{Inverse problem approach in particle digital holography:
  out-of-field particle detection made possible.}}
  {\protect\JournalTitle{Journal of the Optical Society of America. A, Optics,
  image science, and vision}} \textbf{24}, 3708--3716 (2007).

\bibitem{Kempkes2009}
M.~Kempkes, E.~Darakis, T.~Khanam, A.~Rajendran, V.~Kariwala, M.~Mazzotti,
  T.~J. Naughton, and A.~K. Asundi, \enquote{{Three dimensional digital
  holographic profiling of micro-fibers},} {\protect\JournalTitle{Optics
  express}} \textbf{17}, 2938--2943 (2009).

\bibitem{Li2008}
Y.~Li, E.~Perlman, M.~Wan, Y.~Yang, C.~Meneveau, R.~Burns, S.~Chen, A.~Szalay,
  and G.~Eyink, \enquote{{A public turbulence database cluster and applications
  to study Lagrangian evolution of velocity increments in turbulence},}
  {\protect\JournalTitle{Journal of Turbulence}} \textbf{9}, N31 (2008).

\bibitem{Perlman2007}
E.~Perlman, R.~Burns, Y.~Li, and C.~Meneveau, \enquote{{Data exploration of
  turbulence simulations using a database cluster},}
  {\protect\JournalTitle{Proceedings of the 2007 ACM/IEEE Conference on
  Supercomputing (SC '07)}}  (2007).

\bibitem{Yu2012}
H.~Yu, K.~Kanov, E.~Perlman, J.~Graham, E.~Frederix, R.~Burns, A.~Szalay,
  G.~Eyink, and C.~Meneveau, \enquote{{Studying Lagrangian dynamics of
  turbulence using on-demand fluid particle tracking in a public turbulence
  database},} {\protect\JournalTitle{Journal of Turbulence}} \textbf{13}, 1--29
  (2012).

\bibitem{Crocker1996}
J.~C. Crocker and D.~G. Grier, \enquote{{Methods of Digital Video Microscopy
  for Colloidal Studies},} {\protect\JournalTitle{Journal of Colloid and
  Interface Science}} \textbf{179}, 298--310 (1996).

\bibitem{Bedrossian2018}
M.~Bedrossian, M.~El-Kholy, D.~Neamati, and J.~Nadeau, \enquote{{A machine
  learning algorithm for identifying and tracking bacteria in three dimensions
  using Digital Holographic Microscopy},} {\protect\JournalTitle{AIMS
  Biophysics}} \textbf{5}, 36--49 (2018).

\bibitem{Amaro2011}
H.~M. Amaro, A.~C. Guedes, and F.~X. Malcata, \enquote{{Advances and
  perspectives in using microalgae to produce biodiesel},}
  {\protect\JournalTitle{Applied Energy}} \textbf{88}, 3402--3410 (2011).

\bibitem{Chengala2013a}
A.~Chengala, M.~Hondzo, and J.~Sheng, \enquote{{Microalga propels along
  vorticity direction in a shear flow},} {\protect\JournalTitle{Physical Review
  E - Statistical, Nonlinear, and Soft Matter Physics}} \textbf{87}, 1--7
  (2013).

\bibitem{Vigolo2014}
D.~Vigolo, S.~Radl, and H.~A. Stone, \enquote{{Unexpected trapping of particles
  at a T junction},} {\protect\JournalTitle{Proceedings of the National Academy
  of Sciences}} \textbf{111}, 4770--4775 (2014).

\bibitem{Shin2005}
M.~Shin and D.~L. Koch, \enquote{{Rotational and translational dispersion of
  fibres in isotropic turbulent flows},} {\protect\JournalTitle{Journal of
  Fluid Mechanics}} \textbf{540}, 143--173 (2005).

\bibitem{Katz2006}
E.~Katz, A.~L. Yarin, W.~Salalha, and E.~Zussman, \enquote{{Alignment and
  self-assembly of elongated micronsize rods in several flow fields},}
  {\protect\JournalTitle{Journal of Applied Physics}} \textbf{100} (2006).

\bibitem{Parsa2012}
S.~Parsa, E.~Calzavarini, F.~Toschi, and G.~A. Voth, \enquote{{Rotation rate of
  rods in turbulent fluid flow},} {\protect\JournalTitle{Physical Review
  Letters}} \textbf{109}, 1--5 (2012).

\bibitem{Parsa2011}
S.~Parsa, J.~S. Guasto, M.~Kishore, N.~T. Ouellette, J.~P. Gollub, and G.~A.
  Voth, \enquote{{Rotation and alignment of rods in two-dimensional chaotic
  flow},} {\protect\JournalTitle{Physics of Fluids}} \textbf{23}, 043302
  (2011).

\bibitem{Jeffery1922}
G.~B. Jeffery, \enquote{{The Motion of Ellipsoidal Particles Immersed in a
  Viscous Fluid},} {\protect\JournalTitle{Proceedings of the Royal Society A:
  Mathematical, Physical and Engineering Sciences}} \textbf{102}, 161--179
  (1922).

\bibitem{Marcus2014}
G.~G. Marcus, S.~Parsa, S.~Kramel, R.~Ni, and G.~A. Voth,
  \enquote{{Measurements of the solid-body rotation of anisotropic particles in
  3D turbulence},} {\protect\JournalTitle{New Journal of Physics}} \textbf{16}
  (2014).

\end{thebibliography}
\end{document}